\documentstyle[aps,preprint,eqsecnum,tighten]{revtex}

\begin{document}

\draft

\preprint{{\sl November 1994} \hskip 10.7cm SNUTP-94/131}

\title{Energy levels of the soliton--heavy-meson bound states}

\author{Yongseok Oh}

\address{Department of Physics, National Taiwan University,
         Taipei, Taiwan 10764, Republic of China}

\author{Byung-Yoon Park}

\address{Department of Physics, Chungnam National University,
         Daejeon 305-764, Korea}


\maketitle

\begin{abstract}
We investigate the bound states of heavy mesons with finite masses
to a classical soliton solution in the Skyrme model. For a given model
Lagrangian we solve the equations of motion {\em exactly\/} so that the
heavy vector mesons are treated on the same footing as the heavy
pseudoscalar mesons. All the energy levels of higher grand spin states as
well as the ground state are given over a wide range of the heavy meson
masses. We also examine the validity of the approximations used in the
literatures. The recoil effect of finite mass soliton is na\"{\i}vely
estimated.
\end{abstract}



\section{Introduction}
The bound state approach of Callan and Klebanov \cite{CK} has widened
the applicability of the Skyrme model \cite{Sk} up to heavy flavored
baryons. In this approach, heavy baryons are described by bound states
of the soliton of an $SU(2)$ chiral Lagrangian and the heavy mesons
containing the corresponding heavy flavor. This picture originally
introduced for the strange baryons was suggested to be applied for
the heavy baryons containing one or more heavy quarks such as charm
(c) and bottom (b) quarks by Rho, Riska, and Scoccola \cite{RRS,RS}.
The results on the mass spectra \cite{RS} and magnetic moments
\cite{OMRS} for the charmed baryons were found to be strikingly close
to the quark model description.

Although qualitatively successful, when straightforwardly extended
to the heavy flavors, the way of treating the heavy vector mesons
in the traditional bound state approach \cite{SMNR} cannot be
compatible with the heavy quark symmetry \cite{HQS}. As far as the
strange flavor is concerned, on the analogy of the case of $\rho$
and $\pi$, the vector mesons $K^*$ may be integrated out via {\it
an ansatz\/} in favor of a combination of the background and the
pseudoscalar meson field $K$. (See Sec. III for its detailed form.)
However, such an approximation is valid only when the vector meson
mass is much larger than that of the pseudoscalar meson. Furthermore,
the ansatz suppresses the vector meson fields by a factor inversely
proportional to the vector meson mass, while the heavy quark symmetry
implies that the role of the vector mesons becomes as important as
that of the pseudoscalar mesons in the heavy quark mass limit.

In the work of Jenkins, Manohar, and Wise \cite{JMW} followed by a
burst of publications \cite{HBM1,HBM2,HBM3,MOPR}, this problem has
been neatly solved out. There, the bound state approach is applied
to the heavy meson effective Lagrangian \cite{BDW} that explicitly
incorporates the heavy quark symmetry and the chiral symmetry on the
same footing. The resulting baryon mass spectra show the correct
hyperfine splittings consistent with the heavy quark symmetry. It
emphasizes the essential role of the heavy vector mesons in the
binding mechanism. This model is thoroughly studied in Ref.
\cite{OPM3} for higher spin states and is further applied to the
pentaquark exotic baryons \cite{OPM2}.

However, these works are carried out in the limit of both the number
of color $N_c$ and the heavy quark mass $m_Q$ going to infinity,
where the soliton and the heavy mesons are infinitely heavy and so
sit on the top of each other. This na\"{\i}ve picture would definitely
yield larger binding energies to the heavy-meson--soliton system.
Besides, degeneracies and parity doubling in the heavy baryon
spectroscopy of Ref. \cite{OPM3} would be just an artifact originating
from this assumption. In Ref. \cite{OPM1}, the kinetic effects of
finite heavy mesons have been estimated up to $1/m_Q$ order for the
ground state. It is shown that the kinetic effects amounts to about
0.3 GeV in the charm sector, which is not small compared with the
leading order binding energy of 0.8 GeV. Such a correction is expected
to be more serious for the loosely bound pentaquark exotic states with
a binding energy about 0.3 GeV.

In this paper, on our way of investigating the pentaquark exotic
baryons and the heavy baryon excited states in a more realistic way
\cite{OP2}, we generalize the work of Ref. \cite{OPM1} to obtain the
heavy meson (and antiflavored heavy meson) bound states of higher
grand spin and radially excited bound states. For a given model
Lagrangian, we solve the equations of motion {\em exactly\/} without
adopting any approximations. It enables us to investigate the bound
states over a wide range of the heavy meson masses from the
strangeness sector to the bottom sector, and allows us to examine
the validity of the approximations used in the previous works. We
work simply {\em in the soliton-fixed frame\/}, neglecting any recoil
effects due to the finite soliton mass. At the end, a na\"{\i}ve
estimation of the soliton recoil effects is made by replacing the
heavy meson masses by their reduced ones.

This paper is organized as follows. In the next section we briefly
describe our model Lagrangian for completeness. Then, in Sec. III,
we derive the equations of motion for the heavy mesons under the
influence of the static potentials and analyze the symmetries of the
equations. In Sec. IV we discuss the spherical solutions of the free
Proca equations before we give our numerical results for the wave
functions of the bound states in Sec. V where the energy levels of
the soliton--heavy-meson system are presented and discussed. We
summarize the detailed formulas for the grand spin eigenstates in
Appendix A and the equations of motion for the radial functions in
Appendix B.

\section{Model Lagrangian}
We work with a simple model Lagrangian for a system of Goldstone
bosons and the heavy mesons, which possesses the chiral symmetry
explicitly and restores the heavy quark symmetry as the heavy
meson mass goes to infinity.

As for the Goldstone bosons, we adopt the Skyrme model Lagrangian
\cite{Sk}
\begin{mathletters}
\begin{equation}
{\cal L}_{M}^{\text{SM}}
= \frac{f^2_\pi}{4} \mbox{tr}(\partial_\mu U^\dagger \partial^\mu U )
+ \frac{1}{32e^2} \mbox{tr} [U^\dagger \partial_\mu U,
   U^\dagger \partial_\nu U ]^2 ,
\label{Lsk}\end{equation}
where $f_\pi$ is the pion decay constant ($\approx 93$ MeV empirically)
and $U$ is an $SU(2)$ matrix of the chiral field; viz.,
\begin{equation}
U = {\rm e}^{iM/f_\pi},
\end{equation}
with $M$ being a $2\times 2$ matrix of the pion triplet
\begin{equation}
M = \bbox{\tau}\cdot\bbox{\pi} =  \left( \!\! \begin{array}{cc}
\pi^0  \!\!&\!\! \sqrt2\pi^+ \\
\sqrt2\pi^- \!\!&\!\! -\pi^0  \end{array} \!\!\right).
\end{equation}
Here, the chiral $SU(2)_L\times SU(2)_R$ symmetry is realized
nonlinearly under the transformation of $U$:
\begin{equation}
U \longrightarrow L U R^\dagger,
\end{equation}
with $L \in SU(2)_L$ and $R \in SU(2)_R$. By the help of the Skyrme
term with the Skyrme parameter $e$, the Lagrangian
${\cal L}_M^{\text{SM}}$ supports a stable winding-number-1 soliton
solution under the ``hedgehog" configuration
\end{mathletters}
\begin{equation}
U_0({\bf r}) = \exp[i\bbox{\tau}\cdot\hat{{\bf r}} F(r)],
\label{u0}\end{equation}
where the profile of $F(r)$ is subject to the boundary condition,
$F(0) = \pi$ and $F(r)\rightarrow 0$ as $r$ becomes infinity.
The solution carries a finite mass of order $N_c$ and the winding
number is interpreted as the baryon number. The pion decay constant
and the Skyrme parameter are fixed as $f_\pi$=64.5 MeV and $e$=5.45,
respectively, so that the quantized soliton fits the nucleon and
$\Delta$ masses \cite{ANW}.

Now, consider $j^\pi = 0^-$ and $1^-$ heavy mesons with quantum numbers
of a heavy quark $Q$ and a light antiquark $\bar{q}$. We take them as
point-like objects described by the fields $\Phi$ and $\Phi^*_\mu$,
respectively, which form $SU(2)$ anti-doublets. For example, if the
heavy quark is charm flavored, they are the $D$-meson anti-doublets:
\begin{equation}
\Phi = (D^0, D^+) \hskip 5mm\mbox{and}\hskip 5mm
\Phi^* = (D^{*0}, D^{*+}).
\end{equation}
Their conventional free field Lagrangian density is given by
\begin{equation}
{\cal L}^{\text{free}}_\Phi = \partial_\mu \Phi \partial^\mu
 \Phi^\dagger
 - m^2_\Phi \Phi \Phi^\dagger
 - \textstyle\frac12 \Phi^{*\mu\nu} \Phi^{*\dagger}_{\mu\nu}
 + m^2_{\Phi^*} \Phi^{*\mu} \Phi^{*\dagger}_\mu,
\label{lfree} \end{equation}
where $\Phi^*_{\mu\nu}=\partial_\mu \Phi^*_\nu - \partial_\nu
\Phi^*_\mu$ is the field strength tensor of the heavy vector meson
fields $\Phi^*_\mu$ and $m_\Phi^{}$ ($m^{}_{\Phi^*}$) is the mass
of the heavy pseudoscalar (vector) mesons.

In order to construct a chirally invariant Lagrangian containing
$\Phi$, $\Phi_\mu^*$, and their couplings to the Goldstone bosons,
we need to assign a transformation rule with respect to the chiral
symmetry group $SU(2)_L\times SU(2)_R$ to the heavy meson fields.
For this end, we introduce
\begin{equation}
\xi=U^{\frac12},
\label{xi}\end{equation}
which transforms under the $SU(2)_L\times SU(2)_R$ as
\begin{equation}
\xi \rightarrow \xi^\prime = L \xi \vartheta^\dagger
= \vartheta \xi R^\dagger,
\label{Xct} \end{equation}
where $\vartheta$ is a local unitary matrix depending on $L$, $R$,
and the Goldstone fields $M(x)$. Then, we can construct a vector
field $V_\mu$ and an axial vector field $A_\mu$ as
\begin{equation} \renewcommand\arraystretch{1.5} \begin{array}{l}
V_\mu=\frac12(\xi^\dagger\partial_\mu\xi
             +\xi\partial_\mu\xi^\dagger),\\
A_\mu=\frac{i}2(\xi^\dagger\partial_\mu\xi
             -\xi\partial_\mu\xi^\dagger).
\end{array} \end{equation}
The vector field $V_\mu$ behaves as a gauge field under the local
chiral transformation while the axial vector field transforms
covariantly; viz.,
\begin{equation} \begin{array}{l}
V_\mu \rightarrow V^\prime_\mu=\vartheta V_\mu\vartheta^\dagger
+\vartheta\partial_\mu\vartheta^\dagger,\\
A_\mu \rightarrow A^\prime_\mu=\vartheta A_\mu\vartheta^\dagger.
\end{array} \end{equation}
In terms of $\vartheta$, the chiral transformations of $\Phi$ and
$\Phi^*$ fields are expressed as
\begin{equation}
\Phi\rightarrow \Phi^\prime = \Phi\vartheta^\dagger
\hskip 5mm \mbox{and} \hskip 5mm
\Phi_\mu^* \rightarrow \Phi_\mu^{*\prime} = \Phi_\mu^* \vartheta^\dagger.
\label{Pct} \end{equation}
Therefore, a covariant derivative can be defined as
\begin{equation} \begin{array}{l}
D_\mu \Phi \equiv
\Phi(\stackrel{\leftarrow}{\partial}_\mu + V^\dagger_\mu), \\
D_\mu \Phi^*_\nu \equiv
\Phi^*_\nu (\stackrel{\leftarrow}{\partial}_\mu + V^\dagger_\mu).
\end{array} \label{cd}\end{equation}
Such a complication as the multiplication of the field
$V_\mu^\dagger$ from the right hand side is due to the anti-doublet
structure of the heavy meson fields $\Phi$ and $\Phi^*_\mu$.

Given the above definitions, one can write down the chirally invariant
Lagrangian for $\Phi$ and $\Phi_\mu^*$ with their couplings to the
Goldstone bosons. Up to the first derivatives acting on the Goldstone
boson fields, it has the form  \cite{Yan}
\begin{equation} \renewcommand\arraystretch{1.75}\begin{array}{l}
{\cal L} = {\cal L}_M^{SM} + D_\mu \Phi D^\mu \Phi^\dagger
  - m^2_\Phi \Phi \Phi^\dagger
- \frac12 \Phi^{*\mu\nu} \Phi^{*\dagger}_{\mu\nu}
+ m^2_{\Phi^*} \Phi^{*\mu} \Phi^{*\dagger}_\mu  \\
\hskip 1cm + f_Q ( \Phi A^\mu \Phi^{*\dagger}_\mu
   + \Phi^*_\mu A^\mu \Phi^\dagger)
 + \frac12 g_Q \varepsilon^{\mu\nu\lambda\rho}
(\Phi^*_{\mu\nu} A_\lambda \Phi^{*\dagger}_\rho
+ \Phi^*_\rho A_\lambda \Phi^{*\dagger}_{\mu\nu}),
\end{array} \label{lagrangian} \end{equation}
where $\varepsilon_{0123} = +1$,
$f_Q$ and $g_Q$ are the $\Phi^* \Phi M$ and $\Phi^* \Phi^* M$ coupling
constants, respectively, and the field strength tensor is now newly
defined in terms of the covariant derivative of Eq. (\ref{cd}) as
\begin{equation}
\Phi^*_{\mu\nu} = D_\mu \Phi^*_\nu - D_\nu \Phi^*_\mu.
\end{equation}

In this work, instead of restricting the heavy meson masses, $m_\Phi$
and $m_{\Phi^*}$, to their experimental values, we will take them as
free parameters given by a formula
\begin{equation} \renewcommand\arraystretch{2} \begin{array}{l}
\displaystyle
m_\Phi = \overline{m}_\Phi - \frac{3\kappa}{4\overline{m}_\Phi}, \\
\displaystyle
m_{\Phi^*} = \overline{m}_\Phi + \frac{\kappa}{4\overline{m}_\Phi},
\end{array} \label{mf}\end{equation}
with the averaged meson mass $\overline{m}_\Phi \equiv \frac14(m_\Phi
+ 3 m_{\Phi^*})$. Given in Fig. 1 are the heavy meson masses evaluated
by the formula (\ref{mf}) with the parameter $\kappa$ fixed by the
charmed meson masses $m_D$ and $m_{D^*}$. One can see that the formula
reproduces the heavy meson masses remarkably well not only in the
bottom sector but also in the strangeness sector.

As for the coupling constants, since there is no experimental data
available for them (except the upper limit $|g_Q|^2<0.5$ estimated
via the $D^*$ decay width and $D^{*+}$ decay branching ratio), we
will adopt the heavy-quark-symmetric relation \cite{Yan}
\begin{equation}
f_Q = 2 {m}_{\Phi^*} g_Q,
\label{fQgQ}\end{equation}
and take the nonrelativistic quark model estimation for $g_Q$,
\begin{equation}
g_Q = -0.75
\end{equation}
over a wide range of heavy meson masses.

\section{Equations of Motion for Heavy Meson Eigenmodes}
What interests us is the heavy meson eigenstates bound to the static
potentials formed by the fields,
\begin{equation} \begin{array}{l}
V^\mu = (V^0, {\bf V}) = \bbox{(}
       0, -i(\bbox{\tau}\times\hat{{\bf r}}) \upsilon(r) \bbox{)}, \\
A^\mu = (A^0, {\bf A}) = \bbox{(}
       0, \frac12 [ a_1(r)\bbox{\tau}
       + a_2(r) \hat{{\bf r}} \bbox{\tau}\cdot\hat{{\bf r}} ] \bbox{)},
\end{array} \label{va0}\end{equation}
with $\upsilon(r) = [\sin^2 (F/2)] /r$, $a_1(r) = (\sin F)/r$,
and $a_2(r) = F^\prime - (\sin F)/r$,
(Hereafter, we will denote the derivative with respect to $r$ and
$t$ by a prime and a dot, respectively. That is, $f^\prime \equiv
df/dr$ and $\dot{f} \equiv \partial_0 f$.)
which are provided by the soliton configuration (\ref{u0}). We shall
work in the soliton-fixed frame, neglecting any recoil effects of
finite mass soliton. In Sec. V, we will roughly estimate such effects.

The equations of motion for the heavy mesons can be read off from the
Lagrangian (\ref{lagrangian}) as
\begin{eqnarray}
&&
D_\mu D^\mu \Phi^\dagger + m^2_\Phi \Phi^\dagger
 = f_Q A^\mu \Phi^{*\dagger}_\mu,
\label{eq1}\\
&&
D_\mu \Phi^{*\mu\nu\dagger} + m^2_{\Phi^*} \Phi^{*\nu\dagger}
  = - f_Q A^\nu \Phi^\dagger + g_Q \varepsilon^{\mu\nu\lambda\rho}
        A_\lambda \Phi^{*\dagger}_{\mu\rho}.
\label{eq2}
\end{eqnarray}
In order to avoid any unnecessary complications originating from the
anti-doublet structure of $\Phi$ and $\Phi^*_\mu$, we will work with
$\Phi^\dagger$ and $\Phi^{*\dagger}_\mu$ instead.

An auxiliary condition corresponding to the Lorentz condition
$\partial^\mu \Phi^*_\mu =0$ for the free Proca equations can be obtained
by taking a covariant divergence of Eq. (\ref{eq2}):
\begin{equation}
m^2_{\Phi^*} D_\nu \Phi^{*\nu\dagger} =
  - D_\nu D_\mu \Phi^{*\mu\nu\dagger}
  - f_Q D_\nu(A^\nu \Phi^\dagger) + g_Q \varepsilon^{\mu\nu\lambda\rho}
  D_\nu ( A_\lambda \Phi^{*\dagger}_{\mu\rho} ) .
\label{lc}\end{equation}
Note that none of the terms in the right hand side vanishes identically
and that the auxiliary condition, thus, {\it cannot\/} be simply reduced
to $D^\mu \Phi^*_\mu = 0$. Thus, different from the free Proca equations
for the spin-1 field, it is very difficult to eliminate, for example,
$\Phi^*_0$ by using Eq. (\ref{lc}) at this level. Actually, such an
elimination is not an indispensable process in solving out the Proca
equations as we shall see in the next section.

In terms of $\Phi^{*\mu} = (\Phi^{*0}, \bbox{\Phi}^*)$ and using
the form of (\ref{va0}), Eqs. (\ref{eq1}-\ref{lc}) can be rewritten
explicitly as
\begin{eqnarray}
&&
\ddot{\Phi}^\dagger - {\bf D}^2 \Phi^\dagger + m^2_\Phi \Phi^\dagger
   = - f_Q {\bf A} \cdot \bbox{\Phi}^{*\dagger},
\label{eqp}\\
&&
({\bf D}^2 - m^2_{\Phi^*}) \Phi^{*\dagger}_0
   = {\bf D}\cdot\dot{\bbox{\Phi}}{}^{*\dagger}
   - 2g_Q {\bf A}\cdot({\bf D}\times \bbox{\Phi}^{*\dagger}),
\label{eqp*0} \\
&&
\ddot{\bbox{\Phi}}{}^{*\dagger} + {\bf D}\times({\bf D}\times
  \bbox{\Phi}^{*\dagger}) + m^2_{\Phi^*} \bbox{\Phi}^{*\dagger}
   = {\bf D} \dot{\Phi}^{*\dagger}_0 - f_Q {\bf A} \Phi^\dagger
   - 2g_Q {\bf A}\times (\dot{\bbox{\Phi}}{}^{*\dagger}
   - {\bf D} \Phi^{*\dagger}_0),
\label{eqvp*} \\
&& \renewcommand\arraystretch{1.5} \begin{array}{l}
\displaystyle
 m^2_{\Phi^*} (\dot\Phi^{*\dagger}_0 - {\bf D}\cdot\bbox{\Phi}^{*\dagger})
- {\bf D}\cdot [ {\bf D}\times({\bf D}\times\bbox{\Phi}^{*\dagger}) ] \\
\displaystyle \hskip 6mm =  f_Q {\bf D}\cdot({\bf A}\Phi^\dagger)
+ 2g_Q [ {\bf D}\cdot{\bf A}\times(\dot{\bbox{\Phi}}^{*\dagger}
 - {\bf D}\Phi^{*\dagger}_0)
+ {\bf A}\cdot{\bf D}\times \dot{\bbox{\Phi}}{}^{*\dagger} ],
\end{array} \label{ac}
\end{eqnarray}
with ${\bf D} \equiv - \bbox{\nabla} + {\bf V}$.

To solve them, we need to know the symmetries of the equations of
motion. First of all, they are invariant under parity operations under
which the heavy meson fields and the static soliton field transform as
\begin{equation} \begin{array}{c}
\Phi({\bf r},t) \longrightarrow - \Phi(-{\bf r},t), \\
\Phi_0^*({\bf r},t) \longrightarrow + \Phi^*_0(-{\bf r},t), \hskip 5mm
\bbox{\Phi}^*({\bf r},t)  \longrightarrow  -\bbox{\Phi}^*(-{\bf r},t), \\
U({\bf r},t) \longrightarrow U^\dagger(-{\bf r},t), \hskip 5mm
({\bf V}, {\bf A} \longrightarrow -{\bf V}, +{\bf A}),
\end{array} \end{equation}
where we have used that the heavy mesons have negative intrinsic parity.
Next, due to the correlation of the isospin and angular momentum in the
hedgehog configuration (\ref{u0}) and consequently in the static fields
(\ref{va0}), the equations are only invariant under the ``grand spin"
rotation generated by the operator
\begin{equation}
{\bf K} = {\bf J} + {\bf I} = {\bf L} + {\bf S} + {\bf I},
\end{equation}
where ${\bf L}$, ${\bf S}$, and ${\bf I}$, respectively, denote the
orbital angular momentum, spin, and isospin operator of the heavy mesons.
(See Appendix A for their explicit forms and the corresponding eigenbases.)
Thus, the eigenstates are classified by the grand spin quantum numbers
$(k,k_3)$ and the parity $\pi$.

For a given grand spin $(k,k_3)$ with parity $\pi=(-1)^{k\pm 1/2}$,
the general wavefunction of an energy eigenmode can be written as
\begin{equation} \renewcommand\arraystretch{1.5}\begin{array}{l}
\displaystyle
\Phi^{\dagger}({\bf r},t) = e^{+i\omega t} \varphi(r)
{\cal Y}^{(\pm)}_{kk_3}(\hat {\bf r}), \\
\displaystyle
\Phi^{*\dagger}_0({\bf r},t) = e^{+i\omega t} i \varphi^*_0(r)
{\cal Y}^{(\mp)}_{kk_3}(\hat {\bf r}), \\
\bbox{\Phi}^{*\dagger}({\bf r},t) = e^{+i\omega t}
\left[ \varphi^*_1(r) \,
\hat{{\bf r}} {\cal Y}^{(\mp)}_{kk_3}(\hat {\bf r})
+ \varphi^*_2(r) \, {\bf L} {\cal Y}^{(\pm)}_{kk_3}(\hat {\bf r})
+ \varphi^*_3(r) \,
 {\bf G} {\cal Y}^{(\mp)}_{kk_3}(\hat {\bf r}) \right],
\end{array} \label{wf}\end{equation}
with five \cite{fn1} radial functions $\varphi(r)$ and
$\varphi^*_\alpha(r)$ $(\alpha=0,1,2,3)$. Here, ${\bf G}$ is an
operator defined as ${\bf G}\equiv -i(\hat{{\bf r}}\times {\bf L})$,
and ${\cal Y}^{(\pm)}_{kk_3}(\hat{{\bf r}})$ are the spinor spherical
harmonics obtained by combining the orbital angular momentum
eigenstates and the isospin eigenstates. (See Appendix A for further
details.) Note the anomalous sign convention of the energy in the
exponent for the time evolution of the eigenmodes. It is due to the
fact that we are working with $\Phi^\dagger$ and $\Phi^{*\dagger}_\mu$
instead of $\Phi$ and $\Phi^*_\mu$; thus, the eigenenergy $\omega$ is
positive for the bound states of heavy mesons and is negative for the
``antiflavored" heavy mesons of $\bar{Q}q$ structure. Substituting
Eq. (\ref{wf}) into Eqs. (\ref{eqp}-\ref{eqvp*}), we can obtain the
equations of motion for the radial wavefunctions. Their explicit forms
are given in Appendix B.

In the literatures, in order to reduce the complexity of the equations
of motion, the solutions have been found approximately by adopting
proper ans\"{\it a\/}tze. In case of sufficiently heavy mesons, one may
drop the $1/m_Q$ and higher order terms in Eq. (\ref{lc}) which leads
us to a simple auxiliary condition \cite{OPM1}:
\begin{equation}
 D^\mu \Phi^*_\mu = 0 \hskip 1cm
\mbox{(Ansatz I)}.
\label{ap1}\end{equation}
It enables one to eliminate $\Phi^*_0$ easily in favor of the other
three fields $\bbox{\Phi}^*$. In order to be consistent, one should
also drop the higher order terms in the equations of motion for
$\varphi^*_{1,2,3}$.

On the other hand, in the limit of light mesons one may use the fact
that the vector mesons are {\it much heavier\/} than the pseudoscalar
mesons. Then, the most dominant terms in the Lagrangian are those for
the vector meson mass and the $\Phi\Phi^*M$ interactions with the
coupling constant $f_Q$, which reads
\begin{equation} \renewcommand\arraystretch{1.5} \begin{array}{l}
m^2_{\Phi^*} \Phi^{*\mu} \Phi^{*\dagger}_\mu
+ f_Q (\Phi A^\mu \Phi^{*\dagger}_\mu + \Phi^*_\mu A^\mu \Phi^\dagger) \\
\hskip 1cm   = m^2_{\Phi^*}
\left\{ \Phi^{*\mu} + \varepsilon \Phi A^\mu \right\}
\left\{ \Phi^{*\dagger}_\mu + \varepsilon A_\mu \Phi^\dagger\right\}
- 4 g^2_Q \Phi A^\mu A_\mu \Phi^\dagger,
\end{array} \label{mt} \end{equation}
with $\varepsilon$ abbreviating $2g_Q/m_{\Phi^*}$. Here, we have used
the relation (\ref{fQgQ}) for the coupling constant $f_Q$. Equation
(\ref{mt}) suggests us another ansatz for the vector mesons \cite{SMNR}:
\begin{equation}
\Phi^{*\dagger}_\mu = - \frac{2g_Q}{m_{\Phi^*}} A_\mu \Phi^\dagger
\hskip 1cm
\mbox{(Ansatz II)}.
\label{ap2}\end{equation}
In terms of the radial functions, it reads
\begin{equation}
\textstyle
\varphi^*_0(r) = 0, \hskip 5mm
\varphi^*_1(r) = \frac12 \varepsilon ( a_1 + a_2 ) \varphi(r),
\hskip 5mm
\varphi^*_2(r) = \frac12 \varepsilon \gamma_\mp a_1 \varphi(r)
= - \varphi^*_3(r).
\label{ap2a}\end{equation}
However, the equation of motion for the field $\Phi$ (or $\varphi$)
should not be the one obtained simply by substituting Eq. (\ref{ap2})
or (\ref{ap2a}) into the corresponding equation to eliminate
$\Phi^*_\mu$.  Under this ansatz, the Lagrangian (\ref{lagrangian})
is reduced to that of the pseudoscalar field $\Phi$ only:
\begin{equation} \renewcommand\arraystretch{1.5} \begin{array}{l}
{\cal L}_\Phi = D_\mu \Phi D^\mu \Phi^\dagger
- m^2_\Phi \Phi \Phi^\dagger
- \varepsilon^2 D_\mu \Phi A_\nu
(A^\nu D^\mu \Phi^\dagger - A^\mu D^\nu \Phi^\dagger) \\
\hskip 2cm - 4 g^2_Q \Phi A_\mu A^\mu \Phi^\dagger
+ g_Q \varepsilon^2 \varepsilon^{\mu\nu\lambda\rho}
  (D_\mu \Phi A_\nu A_\lambda A_\rho \Phi^\dagger
   + \Phi A_\rho A_\lambda A_\nu D_\mu \Phi^\dagger),
\end{array} \end{equation}
which yields the equation of motion for $\Phi$ as
\begin{equation} \renewcommand\arraystretch{1.5} \begin{array}{c}
(1 + \varepsilon^2 {\bf A}\cdot{\bf A}) \ddot{\Phi}^\dagger
+ 2g_Q \varepsilon^2 {\bf A}\cdot({\bf A}\times{\bf A})\dot{\Phi}^\dagger
- {\bf D}^2 \Phi^\dagger
 + (m^2_\Phi - 4g_Q^2 {\bf A}\cdot{\bf A})\Phi^\dagger \\
- \varepsilon^2 ({\bf A}\times{\bf D})\cdot ({\bf A}\times{\bf D})
\Phi^\dagger
=0.
\end{array} \end{equation}
Substituting
$\Phi^\dagger({\bf r},t) = e^{i\omega t} \varphi(r) {\cal Y}^{\pm}
(\hat {\bf r})$, we can obtain the equation of motion for $\varphi(r)$
as
\begin{equation}
h(r) \varphi^{\prime\prime} + \tilde{h}^\prime \varphi^\prime
+[f(r) \omega^2 + \lambda(r) \omega - m_\Phi^2 - V_{\text{eff}}(r)]
\varphi = 0,
\label{peq}
\end{equation}
where
\begin{equation} \renewcommand\arraystretch{1.5} \begin{array}{l}
h(r) = 1 + \frac12\varepsilon^2 a_1^2,
\hskip 5mm
\tilde{h}^\prime(r) = {\displaystyle\frac{1}{r^2}} [r^2 h(r) ]^\prime,
\\
f(r) = 1 + \frac14 \varepsilon^2[(a_1+a_2)^2 + 2 a_1^2], \\
\lambda(r) = \frac32 \varepsilon^2 g_Q a_1^2 (a_1+a_2), \\
V_{\text{eff}}(r) = \displaystyle \frac{\lambda_\pm}{r^2}
 + \frac{2\mu_\pm}{r}
\upsilon + 2 \upsilon^2 - g^2_Q [(a_1+a_2)^2 + 2 a_1^2] \\
\hskip 1.2cm + \frac14\varepsilon^2 \displaystyle \left\{
\left(\frac{\mu_\pm}{r}+2 \upsilon\right)^2 a_1^2
 + a_1 (a_1'+a_2') \left(\frac{\mu_\pm}{r}+2 \upsilon\right)
 + (a_1+a_2)^2 \left(\textstyle\frac32 a_1^2 + \displaystyle
\frac{\lambda_\mp + \mu_\mp}{r^2} \right) \right\}.
\end{array} \end{equation}
The similarity of the equation to those of Refs. \cite{CK,SMNR} is
remarkable. Those $\varepsilon$-dependent terms play an important
role in obtaining the bound states in the light meson mass limit.
Note that the factor $\varepsilon$ is not necessarily small for the
Ansatz II to be a good approximation \cite{fn2}. In the strangeness
sector, for example, we have $\varepsilon \sim -0.85$
[in $(ef_\pi)^{-1}$ unit].

However, as the mass of the involving heavy flavor increases, the
mass of the pseudoscalar mesons becomes comparable to that of the
vector mesons, and Ansatz II cannot be expected to work well. Note
that it suppresses the role of the vector mesons by the factor
$\varepsilon$, while the heavy quark symmetry implies that they
are not distinguishable from the pseudoscalars as far as the low
energy strong interactions are concerned. In Sec. V, we will give
the exact solutions with the approximated ones using (\ref{ap1})
and (\ref{ap2}).

\section{Solutions to Free Equations}
Before proceeding, we digress for a while to get an insight from
the {\em spherical wave solutions\/} to the {\em free\/} equations,
which are obtained by inserting (\ref{wf}) into
${\cal L}^{\text{free}}$ of (\ref{lfree}):
\begin{eqnarray}
&& \varphi^{\prime\prime} + \frac2r \varphi^\prime
 - (m^2_\Phi - \omega^2 + \frac{\lambda_\pm}{r^2}) \varphi = 0,
\label{feq1} \\ 
&& \varphi^{*\prime\prime}_0 + \frac2r \varphi^{*\prime}_0
   - (m^2_{\Phi^*} + \frac{\lambda_\mp}{r^2})\varphi^*_0
 = - \omega [ (\varphi^{*\prime}_1 + \frac2r \varphi^*_1)
   - \frac{\lambda_\mp}{r} \varphi^*_3 ],
\label{feq2} \\ 
&& (m^2_{\Phi^*} - \omega^2 + \frac{\lambda_\mp}{r^2}) \varphi^*_1
  = \omega \varphi^{*\prime}_0 + \frac{\lambda_\mp}{r}
   (\varphi^{*\prime}_3 + \frac1r \varphi^*_3),
\label{feq3} \\ 
&&\varphi^{*\prime\prime}_2 + \frac2r \varphi^{*\prime}_2
 - (m^2_{\Phi^*} - \omega^2 + \frac{\lambda_\pm}{r^2}) \varphi^*_2 = 0,
\label{feq4} \\ 
&&\varphi^{*\prime\prime}_3 + \frac2r \varphi^{*\prime}_3
   - ( m^2_{\Phi^*} - \omega^2 ) \varphi^*_3
   = \frac1r \varphi^{*\prime}_1 - \frac1r \omega\varphi^*_0,
\label{feq5} 
\end{eqnarray}
and the Lorentz condition
\begin{equation}
\omega \varphi^*_0
  = (\varphi^{*\prime}_1 + \frac2r \varphi^*_1)
    - \frac{\lambda_\mp}{r}\varphi^*_3.
\label{fac} \end{equation}
For a comparison, we are insisting to find the solutions in the form
of Eq. (\ref{wf}). In case of free fields, $\varphi(r)$ and
$\varphi^*_2(r)$ are completely decoupled to the other radial
functions and Eqs. (\ref{feq1}) and (\ref{feq4}) lead to the
solutions as
\begin{equation}
\varphi(r) = j_{\ell_\pm}^{}(qr)  \hskip 5mm \mbox{and} \hskip 5mm
\varphi^*_2(r) = j_{\ell_\pm}^{}(q^*r),
\end{equation}
where $j^{}_{\ell_\pm}(x)$ is the spherical Bessel function of order
$\ell_\pm(= k \mp 1/2)$ with $q^2 \equiv \omega^2 - m^2_{\Phi}$
and $q^{*2} = \omega^2 - m^2_{\Phi^*}$. That is, we have found
two of {\em four\/} independent solution sets of the equations:
\begin{eqnarray} \renewcommand\arraystretch{1.5}
& \mbox{Solution I :} &
\left\{ \begin{array}{l} \displaystyle
\Phi^\dagger({\bf r},t) = e^{i\omega t} j^{}_{\ell_\pm}(qr)
                   {\cal Y}^{(\pm)}_{k \: m_k} (\hat{{\bf r}}), \\
\Phi^{*\dagger}_\mu({\bf r},t) = 0,
\end{array} \right.
\label{sol1}\\
& \mbox{Solution II :} &
\left\{ \begin{array}{l}
\Phi^\dagger({\bf r},t) = 0, \\
\Phi^{*\dagger}_0({\bf r},t) = 0, \\
\displaystyle
\bbox{\Phi}^{*\dagger}({\bf r},t) = e^{i\omega t} j_{\ell_\pm}^{}(q^*r)
               \,{\bf L} {\cal Y}^{(\pm)}_{k \: m_k}(\hat{{\bf r}}),
\end{array} \right.
\label{sol2}
\end{eqnarray}
up to some proper normalizations. Note that there is no conjugate field
of $\Phi^*_0$ in the ``Solution II," which is consistent with the
Lorentz condition (\ref{fac}).

The other two solution sets can be found by the following way. With
the help of the Lorentz condition (\ref{fac}),  Eqs. (\ref{feq2}),
(\ref{feq4}), and (\ref{feq5}) can be rewritten as
\begin{eqnarray}
&&\varphi^{*\prime\prime}_0 + \frac2r \varphi^{*\prime}_0
 + (q^{*2} - \frac{\lambda_\mp}{r^2})\varphi^*_0 = 0, \\
&&\varphi^{*\prime\prime}_1 + \frac2r \varphi^{*\prime}_1
 + (q^{*2} - \frac{\lambda_\mp+2}{r^2})\varphi^*_1
 + \frac{2\lambda_\mp}{r^2} \varphi^*_3 = 0, \\
&&\varphi^{*\prime\prime}_3 + \frac2r \varphi^{*\prime}_3
 + (q^{*2} - \frac{\lambda_\mp}{r^2})\varphi^*_3
 + \frac{2}{r^2} \varphi^*_1 = 0.
\end{eqnarray}
By diagonalizing the $\varphi^*_1$-$\varphi^*_3$ mixing part as
\begin{equation}
\left(\! \renewcommand\arraystretch{1.2} \begin{array}{cc}
\lambda_\mp + 2 & -2 \\
-2 \lambda_\mp & \lambda_\mp \end{array} \!\right)
 \left(\!\begin{array}{c} \varphi^*_1 \\ \varphi^*_3 \end{array}\!\right)
=  \ell^\prime_\pm (\ell^\prime_\pm + 1)
 \left(\!\begin{array}{c} \varphi^*_1 \\ \varphi^*_3 \end{array}\!\right),
\end{equation}
we can obtain the two remaining independent solutions as
\begin{eqnarray}
&\mbox{Solution III :} & \left\{
\begin{array}{l}
\varphi^*_0(r) = (k+1/2) (q^* / \omega) j_{k+1/2}^{}(q^*r), \\
\varphi^*_1(r) = (k+1/2)j^{}_{k-1/2}(q^*r), \\
\varphi^*_3(r) = j^{}_{k-1/2}(q^*r),
\end{array} \right.
\label{sol3a} \\
&\mbox{Solution IV :} & \left\{
\begin{array}{l}
\varphi^*_0(r) = (k+3/2) (q^* / \omega) j_{k+1/2}^{}(q^*r), \\
\varphi^*_1(r) = (k+3/2)j^{}_{k+3/2}(q^*r), \\
\varphi^*_3(r) = -j^{}_{k+3/2}(q^*r),
\end{array} \right.
\label{sol4a}
\end{eqnarray}
for the $\pi=(-1)^{k+1/2}$ states and
\begin{eqnarray}
&\mbox{Solution III :} & \left\{
\begin{array}{l}
\varphi^*_0(r) = (k+1/2) (q^* / \omega) j_{k-1/2}^{}(q^*r), \\
\varphi^*_1(r) = (k+1/2)j^{}_{k+1/2}(q^*r), \\
\varphi^*_3(r) = -j^{}_{k+1/2}(q^*r),
\end{array} \right.
\label{sol3b} \\
&\mbox{Solution IV :} & \left\{
\begin{array}{l}
\varphi^*_0(r) = (k-1/2) (q^* / \omega) j_{k-1/2}^{}(q^*r), \\
\varphi^*_1(r) = (k-1/2)j^{}_{k-3/2}(q^*r), \\
\varphi^*_3(r) = j^{}_{k-3/2}(q^*r),
\end{array} \right.
\label{sol4b}
\end{eqnarray}
for the $\pi=(-1)^{k-1/2}$ states.

{}From these free solutions we can learn the followings:
\begin{enumerate}
\item
In principle, we can obtain the solutions in the form of Eq. (\ref{wf})
without eliminating one of $\Phi^*_\mu$ at the level of Eqs.
(\ref{eqp}-\ref{ac}). It is enough to keep in mind that, due to the
dependence in field variables (and consequently in the equations of
motion), we can obtain only four independent solutions.

\item
The $\varphi^*_1$-$\varphi^*_3$ mixing is due to the fact that our
grand spin eigenstates, $\hat{{\bf r}} {\cal Y}^{(\pm)}_{k\:m_k}$ and
${\bf G} {\cal Y}^{(\pm)}_{k\:m_k}$, are not the eigenstates of
the orbital angular momentum ${\bf L}^2$. If we have used
$\bbox{\cal Y}^{(i)}_{k\:m_k}$, the conjugate radial functions
would have been decoupled. For example, the solution III of Eq.
(\ref{sol3a}) can be written in a form of
\begin{equation} \begin{array}{l}
\mbox{Solution III : }
 \left\{ \begin{array}{l}
\Phi^\dagger({\bf r},t) = 0, \\
\Phi^{*\dagger}_0({\bf r},t) = i(k+1/2) (q^* / \omega)
    j_{k+1/2}^{}(q^*r){\cal Y}^{(-)}_{k\:m_k}(\hat{{\bf r}}), \\
\bbox{\Phi}^{*\dagger}({\bf r},t) =  j^{}_{k-1/2}(q^*r)
 \left[(k+1/2) \, \hat{{\bf r}} {\cal Y}^{(-)}_{k\:m_k}(\hat{{\bf r}})
+ {\bf G} {\cal Y}^{(-)}_{k\:m_k}(\hat{{\bf r}}) \right]
\end{array} \right. \\
\hskip 46mm = \sqrt{(2k+1)(k+1)} j^{}_{k-1/2}(q^*r)
           \,\bbox{\cal Y}^{(2)}_{k\:m_k}(\hat{{\bf r}}).
\end{array} \end{equation}

\item
The special form of Eq. (\ref{feq3}) and (\ref{beq3}) with no derivative
on $\varphi^*_1$ is also ascribed to our choice of the grand spin
eigenstates. Although one can easily express $\varphi^*_1$ in terms
of the other fields by using the equation
\begin{equation}
\varphi^*_1 = - \frac{\omega \varphi^{*\prime}_0 + \lambda_\mp/r
(\varphi^{*\prime}_3 + \varphi^*_3/r)}
{\omega^2 - m^2_{\Phi^*} - \lambda_\mp/r^2},
\label{phi*1}\end{equation}
it cannot be used in eliminating $\varphi^*_1$ from the equations.
Let's see what happens if one proceeds along this direction.
Substituting Eq. (\ref{phi*1}) and its derivative into Eqs.
(\ref{feq2}) and (\ref{feq5}) leads to
\begin{equation} \renewcommand\arraystretch{1.8} \begin{array}{c}
\displaystyle
\frac{m^2_{\Phi^*} + \lambda_\mp/r^2}{\omega^2 - m^2_{\Phi^*}
  - \lambda_\mp/r^2} \: \varphi^{*\prime\prime}_0
+ \frac{\omega \lambda_\mp/r}{\omega^2 - m^2_{\Phi^*} - \lambda_\mp/r^2}
  \: \varphi^{*\prime\prime}_3 = \tilde p_0 ,\\
\displaystyle
\frac{\omega/r}{\omega^2 - m^2_{\Phi^*} - \lambda_\mp/r^2}
    \: \varphi^{*\prime\prime}_0
+ \frac{\omega^2 - m^2_{\Phi^*} }{\omega^2 - m^2_{\Phi^*} - \lambda_\mp/r^2}
  \: \varphi^{*\prime\prime}_3 = \tilde p_3 ,
\end{array} \label{le} \end{equation}
where $\tilde p_{0,3}$ are the terms without containing the second
order derivatives. In order to obtain $\varphi^{*\prime\prime}_{0,3}$,
we should solve the above linear equations. However, the matrix of
the coefficients is singular at the origin and, thus, one cannot
obtain the correct solutions in this way.
\end{enumerate}

\section{Numerical Solutions and Discussions}
In solving equations numerically, there is a considerable freedom.
We will solve Eqs. (\ref{beq1}) and (\ref{beq3}-\ref{beq6}) by
taking $\varphi$, $\varphi^\prime$, $\varphi^*_0$, $\varphi^*_1$,
$\varphi^*_2$, $\varphi^{*\prime}_2$, $\varphi^*_3$, and
$\varphi^{*\prime}_3$ as {\em eight\/} independent variables.
Note that Eqs. (\ref{beq1}), (\ref{beq4}), and (\ref{beq5}) are the
second order differential equations for $\varphi$, $\varphi^*_2$,
and $\varphi^*_3$, while Eqs. (\ref{beq3}) and (\ref{beq6}) are the
first order ones for $\varphi^*_0$ and $\varphi^*_1$, respectively.
On the other hand, we can take Eqs. (\ref{beq1}), (\ref{beq2}),
(\ref{beq4}), and (\ref{beq5}) as the second order differential
equations for $\varphi$, $\varphi^*_0$, $\varphi^*_2$, and
$\varphi^*_3$ and use Eqs. (\ref{beq3}) and (\ref{beq6}) to eliminate
$\varphi^*_1$ and $\varphi^{*\prime}_1$ appearing in those equations.
One may be tempted to eliminate $\varphi^*_1$ by using only Eq.
(\ref{beq3}), since it does not contain any derivative on $\varphi_1^*$.
However, the numerical program according to such a recipe meets
serious instabilities near the origin due to the singularity in a
linear equation similar to Eq. (\ref{le}).

In Fig. 2, we give the radial wavefunctions for a few eigenstates
(solid lines). For a comparison, we present the approximate solutions
obtained by using Ansatz I (dashed lines) and Ansatz II (dash-dotted
lines) as well. In case of the Ansatz II, $\varphi^*_{1,2}$ are
obtained from the relations (\ref{ap2a}) with the solution of Eq.
(\ref{peq}) substituted for $\varphi$. As we have expected, Ansatz II
works only in the limit case; the approximate radial functions of
Ansatz II are quite close to the exact ones in the strangeness sector
but it becomes a bad approximation in the charm sector. One can see
that the factor $\varepsilon$ of the Ansatz II suppresses the vector
fields too much. On the other hand, Ansatz I works quite well both
in the strangeness sector and in the charm sector. In the bottom
sector, the approximate solutions are indistinguishable from the
exact ones. This observation supports the approximation of Ref.
\cite{OPM1}.

{}From Fig. 3, one may arrive at the same conclusion for the
Ans\"{\it a\/}tze I and II, where we present the eigenenergies of
the $k^\pi$=$\frac12^+$ ground state, $\frac12^+$ radially excited
state, and $\frac12^-$ state as functions of the heavy pseudoscalar
meson mass $m_{\Phi}$.  That is, Ansatz I works well over a rather
wide range of the heavy meson masses while Ansatz II works well only
in the limit cases, say, $m_\Phi < 0.8$ GeV. Therefore, the Ansatz II
can be justified only in the strangeness sector as in Ref. \cite{SMNR}.
Note that the Ansatz I and II could yield {\em lower\/} eigenenergies
than the exact ones. However, it does not contradict our common sense
that the exact solutions should have the lower eigenenergy than the
approximate ones. Through the ans\"{\it a\/}tze we have altered the
interactions of the heavy mesons with the soliton more or less. Note
also that the equations admit one bound state in the strangeness sector
without the higher derivative terms. In case of Ansatz II, the
$\varepsilon$ terms are essential for the existence of the bound state.

Given in Fig. 4 are the {\em binding energies\/} $\omega-m_{\Phi}^{}$
of the bound states as functions of the heavy pseudoscalar meson
mass $m_{\Phi}$. As the heavy meson mass increases, the
$k^\pi=\frac12^+$ ground state energy decreases and it may reach down
to the infinite mass limit $-\frac32 g_Q F'(0) \sim 0.8$ GeV. However,
the kinetic effect turns out to be quite large and the binding energy
does not come to the infinite mass limit value even when the heavy
meson mass is increased up to 10 GeV. Note that, as we have expected
from the heavy quark spin symmetry, the eigenstates of $k = k_\ell
\pm 1/2$ become degenerate as the heavy meson mass increases. A special
care should be taken in the quantization of these degenerate bound
states. \cite{OPM3} In Fig. 4, there appear many bound states for the
heavy mesons in the charm and bottom sector. However, remind that we
have neglected any recoil effects due to the finite soliton mass. At
this point, we may {\em na\"{\i}vely\/} estimate the effects of the
soliton motion by replacing the heavy meson masses with their
corresponding ``reduced" masses defined by
\begin{equation}
\mu_\Phi \equiv \frac{m_\Phi m_{\text{sol}}}{m_\Phi + m_{\text{sol}}},
\label{reduced}
\end{equation}
where $m_{\text{sol}}$ is the soliton mass. With $m_{\text{sol}}=867$
MeV, we have $\mu_D \sim 590$ MeV and $\mu_B \sim 743$ MeV. From
Fig. 4, one can see that only one (or at most two) bound state(s)
could survive when the recoil effects are incorporated. This
correction deserves to be studied further.

As discussed in Refs. \cite{OPM3,OPM2}, the equations admit the bound
state solutions with negative eigenenergies, which can be interpreted
as the bound states of antiflavored heavy mesons. In Fig. 5, we
present the radial functions of a few low-lying bound states in case
of $\bar{Q} = \bar{b}$. It is apparent that the radial functions
spread over a wider range of $r$ than those of $Q=b$ do. The binding
energies of antiflavored heavy mesons are given in Fig. 6 as functions
of the heavy meson mass. In this case, the energy change due to the
recoil effects is comparable to the binding energy and the recoil
effects seem to be crucial for the existence of the bound state.
However, a na\"{\i}ve estimation using (\ref{reduced}) shows that
there still survive a bound state near the threshold, which leaves a
possibility of stable pentaquark baryons.

As a summary, we obtained the energy levels of the soliton--heavy-meson
bound states. The equations of motion are solved exactly in a given model
Lagrangian for the excited states as well as for the ground state. The
calculation was also made for the pentaquark exotic baryons. However, to
obtain the real mass spectrum of heavy baryons, we should go one step
further; i.e., the system should be quantized for describing baryons of
definite spin and isospin quantum numbers. Work in this direction is in
progress and will be reported elsewhere.

\vskip 0.3cm
{\it Note added.}\/  After completion of this paper, we were aware of
recent work of Schechter and Subbaraman \cite{SS94}. In estimating the
kinetic effects of heavy mesons, the authors adopt an interesting
approximation to the equations of motion. Corrections due to finite
soliton mass are roughly calculated in the same way as in this paper.

\acknowledgements
This work was supported in part by the National Science Council of
ROC under Grant No. NSC84-2811-M002-036 and in part by the Korea Science
and Engineering Foundation through the SRC program.

\appendix
\section{Grand Spin Eigenstates}
In this Appendix, we construct the grand spin eigenstates, i.e., the
angular part of the wavefunction (\ref{wf}), by combining the orbital
angular momentum (${\bf L}$), spin (${\bf S}$), and isospin (${\bf I}$)
of the heavy mesons. We first combine the orbital angular momentum and
spin to get the angular momentum (${\bf J}$) eigenstates, and then
combine the isospin \cite{fn3}.

To do this, we first find the spin operator $S_i$ $(i=1,2,3)$ for the
vector mesons and the corresponding eigenstates. The Lagrangian
(\ref{lfree}) is invariant under an infinitesimal Lorentz
transformation
\begin{equation} \renewcommand\arraystretch{1.5} \begin{array}{l}
x^\mu \rightarrow x^{\prime\mu} = x^\mu + \epsilon^\mu{}_{\nu} x^\nu, \\
\Phi(x) \rightarrow \Phi^\prime(x^\prime) = \Phi(x), \\
\Phi^*_\alpha(x) \rightarrow \Phi^{\prime *}_\alpha(x^\prime)
 = \frac12 \epsilon^{\mu\nu}(S_{\mu\nu})_\alpha{}^\beta \Phi^*_\beta(x),
\end{array} \end{equation}
where $\epsilon^{\mu\nu} = -\epsilon^{\nu\mu}$ and
$(S^{\mu\nu})_{\alpha\beta}
 = g^\mu{}_\alpha g^\nu{}_\beta - g^\mu{}_\beta g^\nu{}_\alpha$
with $g_{\mu\nu}=\mbox{diag}(1,-1,-1,-1)$.
It defines conserved angular momentum
\begin{equation} \renewcommand\arraystretch{1.5} \begin{array}{l}
J^i = \frac12 \varepsilon^{ijk} \displaystyle \int\!\! d^3r
{\cal M}_{0jk}, \\
{\cal M}_{0jk} = (x^j {\cal P}^{0k} - x^k {\cal P}^{0j})
  + [\Pi^{*m}(S^{kj})_{mn} \Phi^{*n\dagger}
  + \Phi^{*m}(S^{kj})_{mn} \Pi^{*n\dagger}],
\end{array} \end{equation}
where ${\cal P}^{\mu\nu}$ is the canonical energy-momentum tensor and
$\Pi^{*n}(\equiv \partial{\cal L}^{\text{free}}_{\Phi}/\partial
\dot{\Phi}^*_n)$ is the momentum conjugate to the field $\Phi^*_n$.
Here, the indices run from 1 to 3. The first part corresponds to the
orbital angular momentum ${\bf L}$ and the second to the spin angular
momentum ${\bf S}$ of the vector mesons. The latter can be rewritten
neatly as
\begin{equation}
{\bf S} = \int\! d^3r \;(\bbox{\Pi}^{*} \times \bbox{\Phi}^{*\dagger}
 + \bbox{\Phi}^* \times \bbox{\Pi}^{*\dagger}),
\end{equation}
which defines the corresponding quantum mechanical spin operator
acting on the vector meson wavefunctions as
\begin{equation}
S_1 = i \hat{{\bf e}}_1 \times, \hskip 5mm
S_2 = i \hat{{\bf e}}_2 \times, \hskip 5mm
S_3 = i \hat{{\bf e}}_3 \times,
\end{equation}
where $\hat{{\bf e}}_1$, $\hat{{\bf e}}_2$, and $\hat{{\bf e}}_3$ are
unit vectors along the $x$, $y$, and $z$ axis, respectively. Strictly
speaking, although they satisfy the correct commutation relations
$ [S_i, S_j] = i \varepsilon_{ijk} S_k$, and the square is an
invariant of the group, it is not the relativistic spin operator that
commutes with all the generators of the Lorentz group. It is a good
spin operator only in the rest frame, which is, however, enough for
us to proceed. Actually, the invariance of the equations
(\ref{eqp}-\ref{ac}) under the grand spin rotation can be achieved
only with this spin operator.

Eigenvectors of ${\bf S}^2$ and $S_3$ are found by taking suitable
linear combinations of the unit vectors $\hat{{\bf e}}_1$,
$\hat{{\bf e}}_2$, and $\hat{{\bf e}}_3$ \cite{Edmonds}. Let's define
\begin{equation} \renewcommand\arraystretch{1.5} \begin{array}{l}
\hat{{\bf e}}_{\pm1} = \mp\frac{1}{\sqrt2} (\hat{{\bf e}}_1 \pm i
   \hat{{\bf e}}_2), \\
\hat{{\bf e}}_0 = \hat{{\bf e}}_3,
\end{array}
\end{equation}
which satisfy
\begin{equation} \renewcommand\arraystretch{1.3} \begin{array}{l}
{\bf S}^2 \, \hat{{\bf e}}_q^{} = 2 \, \hat{{\bf e}}_q^{}, \\
S_3 \, \hat{{\bf e}}_q^{} = q \, \hat{{\bf e}}_q^{},
\end{array} \hskip 6mm \mbox{ with $q = \pm 1, 0.$}
\end{equation}

Then, the eigenvectors of ${\bf J}^2$ and $J_3$ for the pseudoscalar
mesons are simply the spherical harmonics $Y_{j m_j}(\hat{{\bf r}})$,
and for the vector mesons the ``vector spherical harmonics"
\begin{equation}
{\bf Y}_{j\: \ell \: m_j}(\hat{{\bf r}})
 = \sum_{m,q} Y_{\ell\: m}(\hat{{\bf r}}) \, \hat{{\bf e}}_q \,
   (\ell \: m \: 1 \: q |j \: m_j),
\label{vsh}\end{equation}
where $(\ell \: m \: 1 \: q|j \: m_j)$ is the Clebsch-Gordan coefficient
of adding the orbital angular momentum and the spin. As a result of
the angular momentum addition rule, there can be three different kinds
of vector spherical harmonics with a given $(j, m_j)$:
\begin{equation} \renewcommand\arraystretch{1.2} \begin{array}{ll}
{\bf Y}_{j \: j \: m_j}(\hat{{\bf r}}) &
         \mbox{with parity $\pi=-(-1)^j$ }, \\
{\bf Y}_{j \: j\pm 1 \: m_j}(\hat{{\bf r}}) &
\mbox{ with parity $\pi=+(-1)^{j}$},
\end{array} \end{equation}
where the parity $\pi$ incorporates the intrinsic parity of the vector
mesons.

The vector spherical harmonics can be generated from the spherical
harmonics $Y_{\ell m}(\hat{{\bf r}})$ by making use of certain
operators: \cite{VMK}
\begin{equation}\renewcommand\arraystretch{1.8} \begin{array}{l}
\displaystyle
{\bf Y}_{j \: j \: m_j}
= \frac{1}{\sqrt{j(j+1)}} \, {\bf L} Y_{j \: m_j}, \\
\displaystyle
{\bf Y}_{j \: j+1 \: m_j}
= \sqrt{ \frac{j}{2j+1} } \frac{1}{\sqrt{j(j+1)}} \,  {\bf G} Y_{j \: m_j}
- \sqrt{ \frac{j+1}{2j+1} } \,  \hat{{\bf r}} Y_{j \: m_j}, \\
\displaystyle
{\bf Y}_{j \: j-1 \: m_j}
= \sqrt{ \frac{j+1}{2j+1} } \frac{1}{\sqrt{j(j+1)}} \,  {\bf G} Y_{j \: m_j}
+ \sqrt{ \frac{j}{2j+1} } \,  \hat{{\bf r}} Y_{j \: m_j},
\end{array} \label{vsh1}\end{equation}
where ${\bf G} = -i (\hat{{\bf r}}\times {\bf L})$.
It enables us to carry out somewhat tedious calculations
involving vector spherical harmonics through elementary
vector algebra without referring Clebsch-Gordan coefficients.

To complete the job, we combine the isospin to these angular momentum
eigenstates. The quantum mechanical isospin operator acting on the
{\em isodoublet\/} (not anti-doublet) structure $\Phi^{\dagger}$ and
$\Phi^{*\dagger}_\mu$ is given by the Pauli matrices as
\begin{equation}
{\bf I} = \textstyle \frac12\bbox{\tau}.
\end{equation}
We denote the eigenstates of ${\bf I}^2$ and $I_3$ as
\begin{equation}
\chi_{+\frac12} = \left(\! \begin{array}{c} 1 \\ 0 \end{array} \!\right),
\hskip 7mm
\chi_{-\frac12} = \left(\! \begin{array}{c} 0 \\ 1 \end{array} \!\right),
\end{equation}
which have the properties
${\bf I}^2 \chi_{\pm\frac12} = \frac34 \chi_{\pm\frac12}$
and $I_3 \chi_{\pm\frac12} = \pm\frac12\chi_{\pm\frac12}$.

Now, for the pseudoscalar mesons, the grand spin eigenstates are
obtained simply by the spinor spherical harmonics:
\begin{equation}\renewcommand\arraystretch{1.8} \begin{array}{l}
\hskip -1cm
\mbox{(i) for $k = \ell + 1/2$ and $\pi = (-1)^{k+1/2}$, } \\
\displaystyle
{\cal Y}^{(+)}_{k \: m_k}(\hat{{\bf r}}) =
+ \sqrt{ \frac{k+m_k}{2k} } \, Y_{\ell \: m_k-1/2} \; \chi_{+\frac12}
+ \sqrt{ \frac{k-m_k}{2k} } \, Y_{\ell \: m_k+1/2} \; \chi_{-\frac12}, \\
\hskip -1cm
\mbox{(ii) for $k = \ell - 1/2$ and $\pi = (-1)^{k-1/2}$, } \\
\displaystyle
{\cal Y}^{(-)}_{k \: m_k}(\hat{{\bf r}}) =
- \sqrt{ \frac{k-m_k+1}{2(k+1)} } \, Y_{\ell \: m_k-1/2} \; \chi_{+\frac12}
+ \sqrt{ \frac{k+m_k+1}{2(k+1)} } \, Y_{\ell \: m_k+1/2} \; \chi_{-\frac12},
\end{array} \end{equation}
which are related with each other by
\begin{equation}
-(\bbox{\tau}\cdot\hat{{\bf r}}) \: {\cal Y}^{(\pm)}_{k \: m_k}
 = {\cal Y}^{(\mp)}_{k \: m_k}.
\end{equation}

As for the grand spin eigenstates of the vector mesons, coupling of
the isospin eigenstates to the vector spherical harmonics (\ref{vsh})
leads to six different grand spin eigenstates
$\bbox{\cal Y}^{(i)}_{k \: k_3}$ for a given $(k,k_3)$:
\begin{equation}
\bbox{\cal Y}^{(i)}_{k \: m_k}(\hat{{\bf r}})
= \sum_{m_j q} {\bf Y}_{j_i \ell_i m_j}(\hat{{\bf r}})
\,\chi_{q}^{} \,\textstyle (j_i \: m_j \: \frac12 \: q | k \: m_k),
\end{equation}
with $k = j_i \pm 1/2$ and $j_i = \ell_i, \ell_i \pm 1$. We list
possible $j_i$ and $\ell_i$ in Table \ref{kb}. By the help of Eq.
(\ref{vsh1}), we can rewrite them in terms of $\hat{{\bf r}}
{\cal Y}^{(\pm)}_{k\:m_k}$, ${\bf L} {\cal Y}^{(\pm)}_{k\:m_k}$,
and ${\bf G} {\cal Y}^{(\pm)}_{k\:m_k}$, for example,
\begin{equation}\renewcommand\arraystretch{1.8} \begin{array}{l}
\displaystyle
\bbox{\cal Y}^{(1)}_{k \: m_k}
= \frac{1}{\sqrt{j(j+1)}} \, {\bf L} {\cal Y}^{(+)}_{k \: m_k}, \\
\displaystyle
\bbox{\cal Y}^{(5)}_{k \: m_k}
= \sqrt{ \frac{j+1}{2j+1} } \frac{1}{\sqrt{j(j+1)}}
        \, {\bf G} {\cal Y}^{(+)}_{k \: m_k}
+ \sqrt{ \frac{j}{2j+1} } \, \hat{{\bf r}} {\cal Y}^{(+)}_{k \: m_k}, \\
\displaystyle
\bbox{\cal Y}^{(6)}_{k \: m_k}
= \sqrt{ \frac{j}{2j+1} } \frac{1}{\sqrt{j(j+1)}}
        \, {\bf G} {\cal Y}^{(+)}_{k \: m_k}
- \sqrt{ \frac{j+1}{2j+1} } \, \hat{{\bf r}} {\cal Y}^{(+)}_{k \: m_k},
\end{array} \end{equation}
with $j=k+1/2$. For a practical convenience, we take $\hat{{\bf r}}
{\cal Y}^{(\pm)}_{k\:m_k}$, ${\bf L} {\cal Y}^{(\pm)}_{k\:m_k}$,
and ${\bf G} {\cal Y}^{(\pm)}_{k\:m_k}$ as six independent grand spin
eigenstates instead of $\bbox{\cal Y}^{(i)}_{k\:m_k}$.

The general solution to Eqs. (\ref{eqp}-\ref{ac}) for a given
$(k, m_k)$ can now be written as
\begin{equation} \renewcommand\arraystretch{2} \begin{array}{l}
\displaystyle
\Phi^\dagger({\bf r},t) = e^{+i \omega t}
 \left[ \varphi^{(+)}(r) {\cal Y}^{(+)}_{k \: m_k}(\hat{{\bf r}})
 + \varphi^{(-)}(r) {\cal Y}^{(-)}_{k \: m_k}(\hat{{\bf r}}) \right], \\
\displaystyle
\Phi^{*\dagger}_0({\bf r},t) = i e^{+i \omega t}
 \left[ \varphi^{*(-)}_0(r) {\cal Y}^{(+)}_{k \: m_k}(\hat{{\bf r}})
 + \varphi^{*(+)}_0(r) {\cal Y}^{(-)}_{k \: m_k}(\hat{{\bf r}}) \right], \\
\displaystyle
\bbox{\Phi}^{*\dagger}({\bf r},t) = e^{+i \omega t}
\left[ \varphi^{*(+)}_1(r) \,\hat{{\bf r}}{\cal Y}^{(-)}_{k \: m_k}
  (\hat{{\bf r}})
 + \varphi^{*(+)}_2(r) \,{\bf L}{\cal Y}^{(+)}_{k \: m_k}(\hat{{\bf r}})
 + \varphi^{*(+)}_3(r) \,{\bf G}{\cal Y}^{(-)}_{k \: m_k}(\hat{{\bf r}})
\right. \\
 \hskip 2.5cm \displaystyle
 + \left. \varphi^{*(-)}_1(r) \,\hat{{\bf r}}{\cal Y}^{(+)}_{k \: m_k}
    (\hat{{\bf r}})
 + \varphi^{*(-)}_2(r) \,{\bf L}{\cal Y}^{(-)}_{k \: m_k}(\hat{{\bf r}})
 + \varphi^{*(-)}_3(r) \,{\bf G}{\cal Y}^{(+)}_{k \: m_k}(\hat{{\bf r}})
\right]. \\
\end{array} \end{equation}
We may decompose them into two solution sets of definite parity as
in Eq. (\ref{wf}).

\section{Equations of Motion for Radial Functions}
In this Appendix, we give the explicit forms of the equations of
motion for the radial functions $\varphi(r)$ and $\varphi^*_\alpha(r)$
$(\alpha=0,1,2,3)$. Substitution of Eq. (\ref{wf}) into the equations
of motion (\ref{eqp}-\ref{ac}) leads us to the following coupled
differential equations,
\begin{eqnarray}
&&
\varphi'' + \frac{2}{r} \varphi' + \left[ \omega^2 - m_\Phi^2 -
  \left( \frac{\lambda_\pm}{r^2} + \frac{2\upsilon}{r} \mu_\pm
  + 2 \upsilon^2 \right) \right] \varphi \nonumber \\
&& \hskip 1cm \textstyle
    = - \frac{1}{2} f_Q (a_1 + a_2) \varphi_1^*
    + \frac12 f_Q a_1 (\mu_\pm \varphi_2^* + \mu_\mp \varphi^*_3),
\label{beq1} \\
&&
{\varphi_0^*}'' + \frac{2}{r} {\varphi^*_0}' - \left[ m_{\Phi^*}^2
  + \left( \frac{\lambda_\mp}{r^2} + \frac{2\upsilon}{r} \mu_\mp
  + 2 \upsilon^2 \right) \right] {\varphi^*_0} \nonumber \\
&& \hskip 1cm
= \omega \left[ - \left( {\varphi^*_1}' + \frac{2}{r} {\varphi^*_1}
  \right) + \mu_\pm \upsilon {\varphi^*_2} + \left(
  \frac{\lambda_\mp}{r} + \mu_\mp \upsilon \right) \varphi_3^* \right]
 \label{beq2} \\
&& \hskip 1.5cm
+ g_Q \left[ (a_1 + a_2) f_1 - a_1 (\mu_\mp f_2 + \mu_\pm f_3 ) \right],
  \nonumber \\
&&
\left[ m_{\Phi^*}^2 - \omega^2 + \left( \frac{\lambda_\mp}{r^2} +
  \frac{2\upsilon}{r} \mu_\mp + 2 \upsilon^2 \right) \right]
  {\varphi^*_1} \nonumber \\
&& \hskip 1cm
= \mu_\pm \upsilon \left( {\varphi_2^*}' + \frac{1}{r} {\varphi_2^*}
  \right) + \left( \frac{\lambda_\mp}{r} + \mu_\mp \upsilon \right)
  \left( {\varphi_3^*}' + \frac{1}{r} {\varphi_3^*} \right)
  + \omega {\varphi_0^*}' \label{beq3} \\
&& \hskip 1.5cm \textstyle
 + \frac12 f_Q ( a_1 + a_2 ) \varphi
   - g_Q a_1 \left( \mu_\pm g_2 + \mu_\mp g_3 \right), \nonumber \\
&&
{\varphi_2^*}'' + \frac{2}{r} {\varphi_2^*}' + \left[ \omega^2
  - m_{\Phi^*}^2 - \left( \frac{\lambda_\pm}{r^2}
  + \frac{2 \upsilon}{r} \mu_\pm + \gamma_\pm \mu_\pm \upsilon^2
  \right) \right] {\varphi_2^*} \nonumber \\
&& \hskip 1cm
= \gamma_\pm \upsilon {\varphi_1^*}' + \gamma_\pm \left( \upsilon'
  + \frac{\upsilon}{r} \right) {\varphi_1^*} + \mu_\mp \upsilon \left(
  \frac{1}{r} + \gamma_\pm \upsilon \right) {\varphi_3^*}
   - \omega \gamma_\pm \upsilon {\varphi_0^*}  \label{beq4} \\
&& \hskip 1.5cm \textstyle+ \frac12 f_Q a_1
  \gamma_\pm \varphi + g_Q \left\{ \gamma_\pm a_1 g_1
  + ( a_1 + a_2 ) \left[ \gamma_\pm g_2 + ( 1 + \gamma_\pm ) g_3
  \right] \right\},  \nonumber \\
&&
{\varphi_3^*}'' + \frac{2}{r} {\varphi_3^*}' + \left[ \omega^2 -
  m_{\Phi^*}^2 + \gamma_\pm \mu_\mp \upsilon^2 \right] {\varphi_3^*}
  \nonumber \\
&& \hskip 1cm
= \left( \frac{1}{r} - \gamma_\pm \upsilon \right) {\varphi_1^*}'
  - \gamma_\pm \left( \upsilon' + \frac{\upsilon}{r} \right)
  {\varphi_1^*} - \mu_\pm \upsilon \left( \frac{1}{r} + \gamma_\pm \upsilon
  \right) {\varphi_2^*} - \omega \left( \frac{1}{r} - \gamma_\pm
  \upsilon \right) {\varphi_0^*} \label{beq5}  \\
&& \hskip 1.5cm \textstyle
- \frac12 f_Q a_1 \gamma_\pm \varphi + g_Q \left\{ - \gamma_\pm
  a_1 g_1 + ( a_1 + a_2 ) \left[ ( 1 - \gamma_\pm ) g_2 - \gamma_\pm
  g_3 \right] \right\}, \nonumber \\
&&
m^2_{\Phi^*} \left\{ \omega \varphi^*_0 - (\varphi^{*\prime}_1
 + \frac2r\varphi^*_1) + \mu_\pm \upsilon \varphi^*_2
 + ( \frac{\lambda_\mp}{r} + \mu_\mp \upsilon) \varphi^*_3 \right\}
\nonumber \\
&& \hskip 1cm
+ 2 \upsilon \left( \frac{1}{r} - \upsilon \right) f_1 - \left( \upsilon' +
\frac{\upsilon}{r} \right) \left[ \mu_\mp f_2 + \mu_\pm f_3
  \right]  \nonumber \\
&& = -\frac{f_Q}{2} \left[ (a_1+a_2)(\varphi^\prime
 + \frac2r \varphi) + (a_1+a_2)^\prime \varphi
+ (\frac{\mu_\mp}{r}+2\upsilon) a_1 \varphi \right] \label{beq6} \\
&& \hskip 1cm
+ 2 g_Q \left[ a_1 \left( \upsilon' + \frac{\upsilon}{r} \right)
+ \upsilon ( a_1 + a_2 ) \left( \frac{1}{r} - \upsilon \right)
\right] \varphi^*_0,
\nonumber
\end{eqnarray}
where $f_i$ and $g_i$ ($i$=1,2,3) are functions defined by
\begin{eqnarray}
{\bf D} \times \bbox{\Phi}^{*\dagger} \equiv
i \left[ f_1 \, \hat{{\bf r}} {\cal Y}^{(\pm)}_{k\:m_k}
+ f_2 \, {\bf L} {\cal Y}^{(\mp)}_{k\:m_k}
+ f_3 \, {\bf G} {\cal Y}^{(\pm)}_{k\:m_k}\right], \\
\dot{\bbox{\Phi}}{}^{*\dagger} - {\bf D} \Phi^{*\dagger}_0 \equiv
i \left[ g_1 \, \hat{{\bf r}} {\cal Y}^{(\mp)}_{k\:m_k}
+ g_2 \, {\bf L} {\cal Y}^{(\pm)}_{k\:m_k}
+ g_3 \, {\bf G} {\cal Y}^{(\mp)}_{k\:m_k}\right],
\end{eqnarray}
and $\lambda_{\pm}$, $\mu_{\pm}$, and $\gamma_\pm$ are the numbers
defined through
\begin{equation} \renewcommand\arraystretch{2} \begin{array}{c}
{\bf L}^2 {\cal Y}^{(\pm)}_{k\:m_k}
 \equiv \lambda_{\pm} {\cal Y}^{(\pm)}_{k \: m_k}, \hskip 5mm
(\bbox{\tau}\cdot{\bf L}) {\cal Y}^{(\pm)}_{k\:m_k}
\equiv \mu_{\pm} {\cal Y}^{(\pm)}_{k\:m_k}, \\
\bbox{\tau} {\cal Y}^{(\pm)}_{k \: m_k} \equiv
-\hat{{\bf r}} {\cal Y}^{(\mp)}_{k \: m_k}
+ \gamma_\pm \, {\bf L}{\cal Y}^{(\pm)}_{k \: m_k}
+ \gamma_\mp \, {\bf G}{\cal Y}^{(\mp)}_{k \: m_k},
\end{array} \end{equation}
which also give $\bbox{\tau} \cdot {\bf G} {\cal Y}^{(\pm)}_{k \: m_k}
 = \mu_\pm {\cal Y}^{(\mp)}_{k \: m_k}$.
Explicitly, $f_i$ and $g_i$ ($i$=1,2,3) are given in terms of
$\varphi^*_\alpha$ ($\alpha$=0,1,2,3) as
\begin{eqnarray}
f_1 &=& - \left[ \left( \frac{\lambda_\pm}{r} + \mu_\pm \upsilon
        \right) {\varphi_2^*} + \mu_\mp \upsilon \varphi_3^* \right],
\nonumber \\
f_2 &=& + \left[ \left( \frac{1}{r} - \gamma_\pm \upsilon \right)
        \varphi_1^* - \left( {\varphi_3^*}' + \frac{1}{r} \varphi_3^*
        \right) \right],  \\
f_3 &=& + \left[ \gamma_\pm \upsilon {\varphi_1^*} - \left(
        {\varphi_2^*}' + \frac{1}{r} {\varphi_2^*} \right) \right],
\nonumber \\
g_1 &=& \omega \varphi_1^* + {\varphi_0^*}', \nonumber \\
g_2 &=& \omega \varphi_2^* + \gamma_\pm \upsilon {\varphi_0^*}, \\
g_3 &=& \omega \varphi_3^* + \left( \frac{1}{r} - \gamma_\pm \upsilon
        \right) \varphi_0^*, \nonumber
\end{eqnarray}
and $\lambda_\pm$, $\mu_\pm$, and $\gamma_\pm$ are written in terms of $k$ as
\begin{equation} \renewcommand\arraystretch{1}
\begin{array}{l}
\lambda_+ = (k-1/2)(k+1/2), \\
\mu_+ = k-1/2, \\
\gamma_+ = \mu_+/\lambda_+ = 1/(k+1/2), \end{array}
\hskip 1.5cm
\begin{array}{l}
\lambda_- = (k+1/2)(k+3/2), \\
\mu_- = -(k+3/2), \\
\gamma_- = \mu_-/\lambda_- = -1/(k+1/2). \end{array}
\end{equation}
One can further derive some useful relations between them:
\begin{equation} \renewcommand\arraystretch{1.5} \begin{array}{c}
\lambda_\pm = \lambda_\mp + 2 \mu_\mp +2, \hskip 5mm
\mu_\pm = -(\mu_\mp+2), \\
\gamma_\pm = -\gamma_\mp, \hskip 5mm
\mu_\pm \gamma_\pm + \mu_\mp \gamma_\mp = 2, \hskip 5mm
\lambda_\pm + \mu_\pm = \lambda_\mp + \mu_\mp .
\end{array} \end{equation}

Near the origin, the equations of motion behave asymptotically as
\begin{eqnarray}
&& \varphi^{\prime\prime} + \frac2r \varphi^\prime
 - \frac{\lambda_\mp}{r^2} \varphi = 0,
\label{asym1} \\ 
&& \begin{array}{l}
\displaystyle
\varphi^{*\prime\prime}_0 + \frac2r \varphi^{*\prime}_0
   - \frac{\lambda_\pm}{r^2} \varphi^*_0
 = \omega \left\{ -(\varphi^{*\prime}_1 + \frac2r \varphi^*_1)
   + \frac{\mu_\pm}{r} \varphi^*_2
   + \frac{\lambda_\mp \!+\! \mu_\mp}{r} \varphi^*_3 \right\} \\
\displaystyle \hskip 2cm
+ \eta \left\{ \frac{1}{r} ( 2 + \mu_\mp ) \varphi^*_1 - \mu_\pm \varphi^*_2
- \frac{\lambda_\pm + 2 \mu_\pm}{r} \varphi^*_2 - \mu_\mp \left(
\varphi_3^{*\prime} + \frac{1}{r} \varphi^*_3 \right) \right\},
\end{array}\label{asym2} \\ 
&& \frac{\lambda_\pm}{r^2} \varphi^*_1
  = \frac{\mu_\pm}{r} (\varphi^{*\prime}_2 + \frac1r \varphi^*_2)
   + \frac{\lambda_\mp \!+\! \mu_\mp}{r}
         (\varphi^{*\prime}_3 + \frac1r\varphi^*_3)
    + \omega \varphi^{*\prime}_0
    - \eta \frac{\mu_\pm}{r} \varphi^*_0,
\label{asym3} \\ 
&& \begin{array}{l}
\displaystyle
\varphi^{*\prime\prime}_2 + \frac2r \varphi^{*\prime}_2
 - \frac{(\lambda_\pm \!+\! \mu_\pm)(1 \!+\! \gamma_\pm)}{r^2}
    \varphi^*_2 \\
\displaystyle \hskip 1cm
 = \frac{\gamma_\pm}{r}  \varphi^{*\prime}_1
  + \frac{\mu_\mp (1\!+\!\gamma_\pm) }{r^2} \varphi^*_3
  - \omega\frac{\gamma_\pm}{r} \varphi^*_0 
  + \eta \left\{ \gamma_\pm \varphi^{*\prime}_0
    - \frac{1}{r}\varphi^*_0 \right\},
\end{array} \label{asym4} \\ 
&& \begin{array}{l}
\displaystyle
\varphi^{*\prime\prime}_3 + \frac2r \varphi^{*\prime}_3
   - \frac{\gamma_\mp \mu_\mp}{r^2}  \varphi^*_3 \\
\displaystyle \hskip 1cm
   = \frac{(1 \!+\! \gamma_\pm)}{r} \varphi^{*\prime}_1
    - \frac{\mu_\pm (1 \!+\! \gamma_\pm)}{r^2} \varphi^*_2
   - \omega \frac{(1 \!+\! \gamma_\mp)}{r} \varphi^*_0
   + \eta \gamma_\pm \varphi^{*\prime}_0,
\end{array} \label{asym5} \\ 
&& \begin{array}{l}
\displaystyle
\left\{ \omega \varphi^*_0
- (\varphi^{*\prime}_1 + \frac{2}{r} \varphi^*_1)
+ \frac{\mu_\pm}{r} \varphi^*_2
+ \frac{(\lambda_\mp \!+\! \mu_\mp)}{r}\varphi^*_3 \right\} \\
\displaystyle \hskip 1cm
= 3\eta \delta_2 \varphi^*_0
- \delta_2 \left\{ \frac{\mu_\pm}{r} \varphi^*_1
      - \left[ \mu_\pm \varphi^{*\prime}_2
      + \frac{(\lambda_\mp \!-\! 2)}{r} \varphi^*_2 \right]
      + \mu_\mp\varphi^{*\prime}_3 \right\}
- \delta_1 \left\{ \varphi^\prime - \frac{\mu_\mp}{r} \varphi \right\} ,
\end{array}\label{asym6} \end{eqnarray}
where we have used that, near the origin, the function
$F(r)$ behaves as $F(r) = \pi + F'(0) r
 + \frac16 F^{\prime\prime\prime}(0) r^3 + \cdots$, which implies
\begin{equation}
v(r) \sim \frac{1}{r} + O(r), \hskip 5mm
a_1(r) \sim - F'(0) + O(r^2), \hskip 5mm
a_2(r) \sim 2 F'(0) + O(r^2).
\end{equation}
In the expressions (\ref{asym1}-\ref{asym6}), the constants $\eta$,
$\delta_1$, and $\delta_2$ are defined by
\begin{equation}
\eta = g_Q F^\prime(0), \hskip 5mm
\delta_1 = \frac{ f_Q F^\prime(0) }{ 2 m^2_{\Phi^*} } = O(1/m_Q),
\hskip 5mm
\delta_2 = \frac{ [ F^{\prime}(0)]^2 }{ 2m^2_{\Phi^*} } = O(1/m_Q^2).
\end{equation}
With $F^\prime(0) \sim 2ef_\pi$ and explicit heavy meson masses, one
can evaluate the constants as $\delta_1 \sim 0.65$, 0.29, 0.11 and
$\delta_2 \sim 0.31$, 0.06, 0.01 in case of $Q = s$, $c$, $b$,
respectively. Note that, due to the vector potential ${\bf V}
[\sim i(\hat{{\bf r}}\times\bbox{\tau})/r$, near the origin]
in the covariant derivative, the singular structure of the equations
(\ref{asym1}-\ref{asym6}) is quite different from that of the free
equations (\ref{feq1}-\ref{fac}). Since these equations are not all
independent, they yield only four independent asymptotic solutions
that are finite at the origin. We list them in Table \ref{as1}, where
$c^i$ and $c^i_\alpha$ ($i=\text{I},\text{II},\text{III}$,
$\alpha=0,1,2,3$) for $\pi = (-1)^{k+1/2}$ states are constants satisfying
\begin{equation} \renewcommand\arraystretch{1.5} \begin{array}{l}
(k+1/2) c^i_1  - (k+3/2) c^i_2 - (k+3/2)^2 c^i_3
 - (\omega - \eta) c^i_0 = 0, \\
(\omega - 3\eta\delta_2) c^i_0
-[(k+5/2) + \delta_2 (k-1/2)] c^i_1
+(k-1/2)[1 - 2\delta_2 (k+3/2)] c^i_2 \\
\hskip 1.5cm
+ (k+3/2)[(k-1/2) + \delta_2 (k+5/2)] c^i_3
  + 2\delta_1 (k+1) c^i = 0.
\end{array} \end{equation}
Since we have only two equations for five unknowns, we need to fix
three of them. For example, we take ($c^{\text{I}}_{1,2,3}=0,
c^{\text{I}} = 1$), ($c^{\text{II}} = c^{\text{II}}_3 =0,
c^{\text{II}}_2 = 1$), and ($c^{\text{III}} = c^{\text{III}}_2=0,
c^{\text{III}}_3 =1$), which yield three solution sets given in
Table \ref{as1}.

On the other hand, at large $r$ where heavy mesons become free from
the interactions with the soliton, the equations of motion
(\ref{beq1}-\ref{beq6}) approach asymptotically to Eqs.
(\ref{feq1}-\ref{fac}). Since we are interested in the bound states
with $\omega^2 < m^2_{\Phi}, m^2_{\Phi^*}$, instead of the spherical
Bessel function $j_{\ell}(x)$ in the free solutions, the asymptotic
solutions are written in terms of the modified spherical Hankel
functions $\tilde k_{\ell}(x)$ satisfying
\begin{equation}
\tilde k^{\prime\prime}_\ell + \frac{2}{x} \tilde k^\prime_\ell
- \left[ 1 + \frac{\ell(\ell\!+\!1)}{x^2} \right] \tilde k_\ell^{} = 0,
\end{equation}
with recurrence relations
\begin{equation} \renewcommand\arraystretch{2} \begin{array}{l}
\tilde k_{\ell-1}^{} (x) - \tilde k_{\ell+1}^{} (x) = [(2\ell+1)/x]
    \tilde k_\ell^{} (x), \\
\ell \tilde k_{\ell-1} (x) + (\ell+1) \tilde k_{\ell+1} (x) =
    (2\ell+1) \tilde k^\prime_\ell (x).
\end{array} \end{equation}
Listed in Table \ref{as2} are four independent asymptotic solutions
which decay exponentially at infinity, where $Q = \sqrt{m^2_{\Phi} -
\omega^2}$ and $Q^* = \sqrt{m^2_{\Phi^*} - \omega^2}$.

\begin{figure}
\caption{Heavy meson masses and the mass formula (\protect\ref{mf}).
The pseudoscalar meson mass $m_\Phi$ and the vector meson mass
$\protect m_{\Phi^*}$ obtained using Eq. (\protect\ref{mf}) are given
by solid and dashed lines, respectively.}
\end{figure}

\begin{figure}
\caption{Radial wavefunctions of the heavy meson bound states in
arbitrary scale. Solid lines are those obtained by solving the
equations exactly, while dashed lines and dash-dotted lines are the
approximate solutions of Ans\"{\it a\/}tze I and II,
respectively.}
\end{figure}

\begin{figure}
\caption{Binding energies, $\omega-m_\Phi$, of $k^\pi$=$\frac12^+$
ground state, $\frac12^+$ excited state, and $\frac12^-$ state. Solid
lines represent the exact solutions and dashed (dash-dotted) lines
correspond to Ansatz I (II).}
\end{figure}

\begin{figure}
\caption{Binding energies, $\omega-m_{\Phi}$, of the bound states with
$k^\pi$ as functions of the heavy meson mass. Solid (dashed) lines
denote the positive (negative) parity states.}
\end{figure}

\begin{figure}
\caption{Radial wavefunctions of the antiflavored heavy meson bound
states in arbitrary scale.}
\end{figure}

\begin{figure}
\caption{Binding energies, $|\omega|-m_{\Phi}$, of the bound states
as functions of the antiflavored heavy meson mass. Solid (dashed)
lines denote the positive (negative) parity states.}
\end{figure}

\begin{table}
\caption{Eight grand spin eigenstates.}
\begin{tabular}{ccccccccc}
 \multicolumn{4}{c}{ $\pi=(-1)^{k+1/2}$ } & \vline &
 \multicolumn{4}{c}{ $\pi=(-1)^{k-1/2}$ } \\
 state & $s$ & $\ell$ & $j$ & \vline & state & $s$ & $\ell$ & $j$ \\
\hline
${\cal Y}^{(+)}_{k \: m_k}$ & 0 & $j$ & $k-1/2$ & \vline &
${\cal Y}^{(-)}_{k \: m_k}$ & 0 & $j$ & $k+1/2$ \\
\hline
$\bbox{\cal Y}^{(1)}_{k \: m_k}$ & 1 & $j$   & $k-1/2$ & \vline &
$\bbox{\cal Y}^{(4)}_{k \: m_k}$ & 1 & $j$   & $k+1/2$ \\
$\bbox{\cal Y}^{(2)}_{k \: m_k}$ & 1 & $j-1$ & $k+1/2$ & \vline &
$\bbox{\cal Y}^{(5)}_{k \: m_k}$ & 1 & $j-1$ & $k-1/2$ \\
$\bbox{\cal Y}^{(3)}_{k \: m_k}$ & 1 & $j+1$ & $k+1/2$ & \vline &
$\bbox{\cal Y}^{(6)}_{k \: m_k}$ & 1 & $j+1$ & $k-1/2$
\end{tabular}
\label{kb} \end{table}

\begin{table}
\caption{Four independent asymptotic solutions near the origin.}
\begin{tabular}{cll}
parity & \multicolumn{1}{c}{ $\pi=(-1)^{k+1/2}$ } &
  \multicolumn{1}{c}{ $\pi=(-1)^{k-1/2}$ } \\
\hline 
Solution I &
$\begin{array}{l}
\varphi \sim r^{k+1/2} + O(r^{k+5/2}) \\
\varphi^*_0 \sim c^{\text{I}}_0 r^{k-1/2} + O(r^{k+3/2}) \\
\varphi^*_{1,2,3} \sim O(r^{k+5/2})
\end{array}$
&
$\begin{array}{l}
\varphi \sim r^{k-1/2} + O(r^{k+3/2}) \\
\varphi^*_0 \sim  O(r^{k+1/2}) \\
\varphi^*_{1,2,3} \sim O(r^{k+3/2})
\end{array}$ \\
\hline 
Solution II &
$\begin{array}{l}
\varphi, \varphi^*_3 \sim  O(r^{k+5/2}) \\
\varphi^*_0 \sim c^{\text{II}}_0 r^{k-1/2} + O(r^{k+3/2}) \\
\varphi^*_1 \sim c^{\text{II}}_1 r^{k+1/2} + O(r^{k+5/2}) \\
\varphi^*_2 \sim r^{k+1/2} + O(r^{k+5/2})
\end{array}$
&
$\begin{array}{l}
\varphi, \varphi^*_3 \sim O(r^{k+3/2}) \\
\varphi^*_0 \sim O(r^{k+1/2}) \\
\varphi^*_1 \sim + r^{k-1/2} + O(r^{k+3/2}) \\
\varphi^*_2 \sim - r^{k-1/2} + O(r^{k+3/2})
\end{array}$ \\
\hline 
Solution III &
$\begin{array}{l}
\varphi, \varphi^*_2 \sim  O(r^{k+5/2}) \\
\varphi^*_0 \sim c^{\text{III}}_0 r^{k-1/2} + O(r^{k+3/2}) \\
\varphi^*_1 \sim c^{\text{III}}_1 r^{k+1/2} + O(r^{k+5/2}) \\
\varphi^*_3 \sim r^{k+1/2} + O(r^{k+5/2})
\end{array}$
&
$\begin{array}{l}
\varphi, \varphi^*_2 \sim O(r^{k+3/2}) \\
\varphi^*_0 \sim O(r^{k+1/2}) \\
\varphi^*_1 \sim (k-1/2) r^{k-1/2} + O(r^{k+3/2}) \\
\varphi^*_3 \sim r^{k-1/2} + O(r^{k+3/2})
\end{array}$ \\
\hline 
Solution IV &
$\begin{array}{l}
\varphi \sim  O(r^{k+1/2}) \\
\varphi^*_0 \sim r^{k-1/2} + O(r^{k+3/2}) \\
\varphi^*_1 \sim \lambda_+ r^{k-3/2} + O(r^{k+1/2}) \\
\varphi^*_2 \sim r^{k-3/2} + O(r^{k+1/2}) \\
\varphi^*_3 \sim \mu_+ r^{k-3/2} + O(r^{k+1/2})
\end{array}$
&
$\begin{array}{l}
\varphi \sim O(r^{k+3/2}) \\
\varphi^*_0 \sim r^{k+1/2} + O(r^{k+5/2}) \\
\varphi^*_1 \sim O(r^{k+3/2}) \\
\varphi^*_2 \sim O(r^{k+3/2}) \\
\varphi^*_3 \sim O(r^{k+3/2})
\end{array}$
\end{tabular}
\label{as1}\end{table}

\begin{table}
\caption{Four independent asymptotic solutions at large $r$.}
\begin{tabular}{cll}
parity & \multicolumn{1}{c}{ $\pi=(-1)^{k+1/2}$ } &
  \multicolumn{1}{c}{ $\pi=(-1)^{k-1/2}$ } \\
\hline 
Solution I &
$\begin{array}{l}
\varphi \sim \tilde k_{k-1/2}(Qr) \\
\varphi^*_{0,1,2,3} \sim 0
\end{array}$
&
$\begin{array}{l}
\varphi \sim \tilde k_{k+1/2}(Qr) \\
\varphi^*_{0,1,2,3} \sim 0
\end{array}$ \\
\hline 
Solution II &
$\begin{array}{l}
\varphi^*_2 \sim \tilde k_{k-1/2}(Q^*r) \\
\varphi, \varphi^*_{0,1,3} \sim  0
\end{array}$
&
$\begin{array}{l}
\varphi^*_2 \sim \tilde k_{k+1/2}(Q^*r) \\
\varphi, \varphi^*_{0,1,3} \sim  0
\end{array}$ \\
\hline 
Solution III &
$\begin{array}{l}
\varphi, \varphi^*_2 \sim  0 \\
\varphi^*_0 \sim (k+1/2)(Q^*/\omega) \tilde k_{k+1/2}(Q^*r) \\
\varphi^*_1 \sim (k+1/2) \tilde k_{k-1/2}(Q^*r) \\
\varphi^*_3 \sim \tilde k_{k-1/2}(Q^*r)
\end{array}$
&
$\begin{array}{l}
\varphi, \varphi^*_2 \sim 0 \\
\varphi^*_0 \sim -(k+1/2) (Q^*/\omega) \tilde k_{k-1/2}(Q^*r) \\
\varphi^*_1 \sim (k+1/2) \tilde k_{k+1/2}(Q^*r) \\
\varphi^*_3 \sim - \tilde k_{k+1/2}(Q^*r)
\end{array}$ \\
\hline 
Solution IV &
$\begin{array}{l}
\varphi, \varphi^*_2 \sim  0 \\
\varphi^*_0 \sim (k+3/2) (Q^*/\omega) \tilde k_{k+1/2}(Q^*r) \\
\varphi^*_1 \sim (k+3/2) \tilde k_{k+3/2}(Q^*r) \\
\varphi^*_3 \sim - k_{k+3/2} \tilde k_{k+3/2}(Q^*r)
\end{array}$
&
$\begin{array}{l}
\varphi, \varphi^*_2 \sim 0  \\
\varphi^*_0 \sim -(k-1/2)(Q^*/\omega) \tilde k_{k-1/2}(Q^*r) \\
\varphi^*_1 \sim  (k-1/2) \tilde k_{k-3/2}(Q^*r) \\
\varphi^*_3 \sim \tilde k_{k-3/2}(Q^*r)
\end{array}$
\end{tabular}
\label{as2}\end{table}


\begin{references}
\bibitem{fn2}
   The most essential point of Ansatz II is that the vector
   mesons are much heavier than the pseudoscalar mesons.
\end{references}
\end{document}